\documentclass[useAMS,usenatbib]{mn2e}

\usepackage{url,times,stfloats,graphicx,amsmath,amsfonts,amssymb,aas_macros,color,epsfig,epstopdf,caption}

\newcommand{\mhalo}{M$_{\rm halo}$}
\newcommand{\mb}{M$_{\rm b}$}
\newcommand{\mstar}{M$_{\rm *}$}
\newcommand{\mhi}{M$_{\rm HI}$}
\newcommand{\lcdm}{$\Lambda$CDM}
\newcommand{\msun}{M$_\odot$}

\newcommand{\vmax}{V$_{\rm max}$}
\newcommand{\vflat}{V$_{\rm flat}$}
\newcommand{\rflat}{R$_{\rm flat}$}
\newcommand{\vlast}{V$_{\rm last}$}
 \newcommand{\rlast}{R$_{\rm last}$}

\newcommand{\wt}{W$_{20}$}
\newcommand{\wf}{W$_{50}$}

\newcommand{\rHI}{R$_{\rm HI}$}
\newcommand{\dHI}{D$_{\rm HI}$}
\newcommand{\vDHI}{V$_{\rm DHI}$}
\newcommand{\dsf}{D$_{\rm 6.4h}$}

\newcommand{\vrot}{V$_{\rm rot}$}

\newcommand{\kms}{km$\,$s$^{-1}$}

\newcommand{\ahf}{\texttt{AHF}}

\newcommand{\hMpc}{{\ifmmode{h^{-1}{\rm Mpc}}\else{$h^{-1}$Mpc}\fi}}
\newcommand{\hkpc}{{\ifmmode{h^{-1}{\rm kpc}}\else{$h^{-1}$kpc}\fi}}
\newcommand{\hMsun}{{\ifmmode{h^{-1}{\rm {M_{\odot}}}}\else{$h^{-1}{\rm{M_{\odot}}}$}\fi}}
\newcommand{\ltsima}{$\; \buildrel < \over \sim \;$}
\newcommand{\gtsima}{$\; \buildrel > \over \sim \;$}
\newcommand{\lsim}{\lower.5ex\hbox{\ltsima}}
\newcommand{\gsim}{\lower.5ex\hbox{\gtsima}}

\def\lesssim{\mathrel{\hbox{\rlap{\hbox{\lower4pt\hbox{$\sim$}}}\hbox{$<$}}}}
\def\gtrsim{\mathrel{\hbox{\rlap{\hbox{\lower4pt\hbox{$\sim$}}}\hbox{$>$}}}}

\newcommand{\beq}{\begin{equation}}
\newcommand{\eeq}{\end{equation}}
\def\beqa{\begin{eqnarray}}
\def\eeqa{\end{eqnarray}}
\def\hMpc{$h^{-1}\,{\rm Mpc}$}
\def\hkpc{$h^{-1}\,{\rm kpc}$}

\def\vmaxDM{V$_{\rm max}^{\rm DM}$}
\def\rmaxDM{R$_{\rm max}^{\rm DM}$}

\def\head{
 \vbox to 0pt{\vss
                   \hbox to 0pt{\hskip 440pt\rm LA-UR-10-07069\hss}
                  \vskip 25pt}}

\title[The different Baryonic Tully Fisher relations]
{The different  baryonic Tully-Fisher relations at low masses.}
\author[C. Brook et al.]
       {Chris B. Brook$^{1,2}$\thanks{E-mail: cbabrook@gmail.com}, Isabel Santos-Santos$^{1}$, Greg Stinson$^{3}$\\
$^{1}$Departamento de F\'isica Te\'orica, Universidad Aut\'onoma de Madrid, 28049 Cantoblanco, Madrid, Spain\\
$^{2}$Astro-UAM, UAM, Unidad Asociada CSIC\\   
$^3$Max-Planck-Institut f\"ur Astronomie, K\"onigstuhl 17, 69117 Heidelberg, Germany\\
}

\begin{document}

\date{Accepted XXXX . Received XXXX; in original form XXXX}

\pagerange{\pageref{firstpage}--\pageref{lastpage}} \pubyear{2010}

\maketitle

\label{firstpage}


\begin{abstract}

We compare the Baryonic Tully-Fisher relation (BTFR) of simulations and observations  of galaxies ranging from dwarfs to spirals, using various measures of rotational velocity \vrot.  
We explore the BTFR when measuring \vrot\ at the flat part of the rotation curve,  \vflat,  at the extent of HI gas,  \vlast, and  using  20\% (\wt) and 50\% (\wf) of the width of HI line profiles. We also compare with the maximum circular velocity of the parent halo, V$_{\rm max}^{\rm DM}$,  within dark matter only simulations. The  different BTFRs  increasingly diverge as galaxy mass decreases. Using \vlast\, one obtains a power law over four orders of magnitude in baryonic mass, with slope similar to the observed BTFR. Measuring \vflat\ gives similar results as \vlast\ when galaxies with rising rotation curves are excluded. However,  higher rotation velocities would be found for low mass galaxies if the cold gas extended far enough for \vrot\ to  reach a maximum. \wt\ gives a similar slope as \vlast\ but with slightly lower values of \vrot\ for low mass galaxies, although this may depend on the extent of the gas in your galaxy sample. \wf\ bends away from these other relations toward low velocities at low masses. By contrast, \vmaxDM\ bends toward high velocities for low mass galaxies, as cold gas does not extend out to the radius at which halos reach \vmaxDM. Our study highlights the need for careful comparisons between observations and models: one needs to be consistent about the  particular method of measuring \vrot, and precise about  the radius at which velocities are measured. 
\end{abstract}

\noindent
\begin{keywords}
 galaxies: evolution - formation - haloes cosmology: theory - dark matter
 \end{keywords}

\section{Introduction} \label{sec:introduction}
Galaxies  follow a tight relation between optical luminosity and the width of the 21\,cm line of neutral hydrogen HI  (Tully \& Fisher 1977).  The 21\,cm line-width is a measure of rotation velocity \vrot\ which reflects total mass within a given radius,  while luminosity is a reflection of stellar mass. Other measures of \vrot\,  have subsequently been used in deriving the Tully-Fisher relation,  using information from the full rotation curves of galaxies  \citep[e.g.][]{verheijen01,mcgaugh05,kuziodenaray06,noordermeer07,yegorova07,oh11}, and luminosity is often converted to stellar masses, making the relation between \vrot\ and stellar mass  more explicit.  


\begin{table*}

\begin{center}

\caption{Properties of the MaGICC galaxies ordered by halo mass. \rHI\ and \vlast\ are  measured at $\dagger\,$N$_{\rm HI}$=1M$_\odot$/pc$^2$ and $\P\,$N$_{\rm HI}$=10$^{19}$cm$^{-2}$}

\begin{tabular}{l l l l l l l l l l l l l l l}

\hline

Name &  \mhalo\  & \mstar\  & M$_{\rm HI}$  & $h_I$  &   R$_{\rm HI}^\dagger$     & R$_{\rm HI}^\P$     &  V$_{\rm last}^\dagger$ &  V$_{\rm last}^\P$

 & R$_{\rm flat}$& V$_{\rm flat}$ & \wt & \wf  & R$_{\rm max}^{\rm DM}$    & V$_{\rm max}^{\rm DM}$    \\

&(\msun)&(\msun)&(\msun)&(kpc)&(kpc)& (kpc)  &(km/s)& (km/s) &  (kpc)&(km/s)&  (km/s)    &(km/s) & (kpc)   &(km/s)\\

\hline

g15784\_MW &1.49$\times$10$^{12}$& 5.67$\times$10$^{10}$& 4.04$\times$10$^{10}$    &3.23 & 31.9 & 46.3 &225.7&212.0   & 26.8 & 218.8 & 500.5 & 404.8  & 54.90& 183.0       \\

g21647\_MW  &8.24$\times$10$^{11}$& 2.51$\times$10$^{10}$& 8.51$\times$10$^9$    & 1.30& 14.5&   33.9& 164.4 &150.7&17.1 & 157.5 & 390.0 & 350.0&  38.15&    161.6    \\

g1536\_MW  &  7.10$\times$10$^{11}$&2.36$\times$10$^{10}$& 1.20$\times$10$^{10}$   &3.46& 33.2 &   42.0 &165.8 &157.6& 8.48&170.8& 386.2 & 349.9& 52.86  &  142.9     \\

g5664\_MW  & 5.39$\times$10$^{11}$&2.74$\times$10$^{10}$& 7.25$\times$10$^9$   &2.34& 17.4 & 42.4  & 162.4&137.6&12.6&  169.0& 399.8 & 368.5&39.15&  124.6      \\

g7124\_MW  & 4.47$\times$10$^{11}$&6.30$\times$10$^9$& 6.99$\times$10$^9$   &2.79 & 15.3  &   30.7  &126.0&120.7& 12.3& 118.2 & 218.6 & 154.8 & 38.15 &     135.1     \\

g15807\_Irr &2.82$\times$10$^{11}$& 1.46$\times$10$^{10}$& 7.66$\times$10$^9$  &1.94&17.0  &    25.9&134.7&123.3 &10.4 &  137.0& 289.4 & 179.2 & 32.99& 101.0     \\

g15784\_Irr &1.70$\times$10$^{11}$& 4.26$\times$10$^9$& 5.11$\times$10$^9$  &2.27 & 12.8    &19.6 & 110.4&104.1 &12.2 &104.8& 210.4 & 183.6 &27.45 &    91.51   \\

g22437\_Irr  &  1.10$\times$10$^{11}$&7.44$\times$10$^8$& 1.87$\times$10$^9$  &1.88 &7.87    & 12.4 & 75.95&77.14& 6.44 &69.44 & 141.0 &104.9& 27.32 &  66.69   \\

g21647\_Irr  &  9.65$\times$10$^{10}$&1.98$\times$10$^8$&  9.55$\times$10$^8$  &1.75& 7.74    & 21.2 & 60.51&61.89& 7.70&56.15& 111.1 &80.00& 19.08 &    80.81    \\

g1536\_Irr & 8.04$\times$10$^{10}$&4.46$\times$10$^8$& 1.25$\times$10$^9$   &1.70& 7.73  & 14.9 & 64.97&68.63& 9.14& 61.66& 94.15 & 65.37 & 26.43 &    71.49   \\

g5664\_Irr  &  5.87$\times$10$^{10}$& 2.36$\times$10$^8$&  7.50$\times$10$^8$   &1.66 &   6.82   &13.6 & 54.85&61.14& 8.92  & 55.27& 105.3& 58.36 & 19.58&    62.31    \\

g7124\_Irr  & 5.23$\times$10$^{10}$&1.32$\times$10$^8$& 7.13$\times$10$^8$   &1.16 &  7.19 & 11.1& 51.53&54.19& 7.37 &48.05  & 81.40 & 52.73 & 19.08&  67.56  \\

g15807\_dIrr &3.04$\times$10$^{10}$&   1.60$\times$10$^{7}$& 1.31$\times$10$^{8}$& 1.26& 2.96  &  7.77 & 29.93&41.86& 8.00   & 41.22&53.70 & 37.36&16.50  &  50.49   \\

g15784\_dIrr  & 1.77$\times$10$^{10}$&8.98$\times$10$^{6}$& 2.12$\times$10$^{7}$&0.73 & 1.48  & 2.06 & 23.26&27.41& 3.92&34.09   & 38.08 & 18.68 &13.73 &  45.75       \\

g22437\_dIrr & 1.19$\times$10$^{10}$&5.05$\times$10$^{5}$& 1.61$\times$10$^{7}$&0.17& 1.23  & 1.59 & 21.92&23.52& 1.08& 20.09&  27.75 & 12.79& 13.66    & 36.17  \\

g1536\_dIrr & 9.69$\times$10$^{9}$& 7.20$\times$10$^{5}$&  1.79$\times$10$^{7}$   &0.17 &1.04 &1.58 &21.72&25.33& 1.58&24.10& 41.96 & 16.81& 13.22& 35.74     \\

\hline

\end{tabular}
\label{tab:sims}
\end{center}
\end{table*}

For low mass galaxies, which become increasingly gas rich \citep{geha06,bradford15},  a  tighter relation is found when \vrot\ is plotted against the total observable baryonic mass \mb, i.e. stellar mass plus cold gas mass \citep{freeman99,mcgaugh00}. 
 As argued in \cite{mcgaugh12}, the gas dominance  in low mass galaxies minimises the importance of the error in stellar mass, allowing relatively accurate measurements of \mb\ when deriving the baryonic Tully-Fisher relation (BTFR).

 However, low mass galaxies are more problematic when it comes to determining \vrot. For high mass galaxies, rotation curves generally rise sharply  \citep{roberts73,rubin78},   and some then drop before flattening at the outer parts  \cite[e.g.][]{sofue01,noordermeer07},   at a velocity which can then be defined as \vflat, a common measure of \vrot\ used in BTFRs.  
 By contrast, rotation curves of low mass galaxies rise more gently   \cite[e.g.][]{lelli13} and although they do start to flatten beyond about 2 disc scale-lengths, they are often still rising at the last measured point \cite[e.g.][]{catinella06,swaters09}. 
To extend the \vflat\ BTFR to low mass galaxies, \cite{stark09}  select galaxies 
with rotation curves that are approximately flat in their outer regions, according to some flatness criteria. 

The low mass end of the BTFR is also problematic when measuring \vrot\ using HI line widths. The discrepancy between measuring the HI line-width at 20 per\,cent (W$_{20}$) and 50 per\,cent (W$_{50}$)  of its peak value is typically 25\,\kms\ \citep{koribalski04,bradford15}, which  becomes an increasing fraction  of the measured rotation velocity at lower masses. 

  Observational studies have shown that the way \vrot\ is measured (\vflat, \vlast, \wt, \wf) will result in different T-F and BTF relations \cite[e.g.][]{verheijen01,noordermeer07,mcgaugh12,bradford16}. Further,  sample selection also plays a  role \citep{sorce16}.
  Understanding the  difference in measured values of \vrot\ has important implications for interpreting the BTFR, which is used as a crucial constraint on theoretical models \cite[e.g.][]{dutton09,trujillogomez11,mcgaugh12,dutton12,brook12a}.
  As recently shown  \citep{brook15b,brookshankar}, 
these differences in measures of \vrot\ also significantly affect the interpretation of the observed velocity function, i.e. the number of galaxies at given \vrot within a given volume \citep{zavala09,obreschkow09,papastergis11,klypin14}. 
 
 In this paper, we explore  the BTFR of the MaGICC suite of simulated galaxies \citep{brook12a,stinson13}, comparing the derived BTFRs when measuring \vrot\  in different ways. We show, in particular, that the different forms of the BTFR become increasingly divergent at low masses.  The paper is organized as follows: Sec~\ref{sims} presents the simulations, describing  initial conditions and  baryonic modelling. The rotation curves and HI line-widths are shown in Sec~\ref{sec:rc}\,\&\,\ref{sec:HI}, respectively. The various derived measures of \vrot\ are then shown in Sec~\ref{sec:vrot}, and the different forms of the BTFR in Sec~\ref{sec:BTFR}, with implications discussed in Sec~\ref{discussion}. 
 
\section{The Simulations}\label{simulations}\label{sims}

We use   `zoom' hydro-dynamical simulations from the MaGICC (Making Galaxies in a Cosmological Context) project \citep{brook12a,stinson13}.  The initial 
power spectrum is derived from the McMaster Unbiased Galaxy Simulations  \citep{stinson10} which use a \lcdm\ cosmology with WMAP3 parameters.  The  simulated galaxies cover a range of merger histories and formation times, but are all isolated. Therefore, whether environment has an effect on  our results cannot be answered in this study.

The simulated galaxies are evolved using the parallel N-body+SPH tree-code \verb,GASOLINE, \citep{wadsley04}, which includes gas hydrodynamics and cooling, star formation, energy feedback and metal enrichment. We describe here the most important implementations (for details see  \citealt{stinson13}).

When gas gets cold and dense, stars form  at  rate $\propto\rho^{1.5}$. Stars feed energy and metals to the surrounding interstellar medium. Energy feedback by supernovae is implemented using the blastwave formalism \citep{stinson06}, releasing $\epsilon_{\rm SN}$$\times$ $10^{51}$ erg of thermal energy,  where we set $\epsilon_{\rm SN}$=1, i..e  we release all available SN energy in thermal form, even though observations indicate that a significant amount of the energy is radiated away. The reason is because thermal energy is radiated away rapidly in the simulations for numerical reasons, due to lack of resolution  (see Stinson  et al. 2013 for a full discussion).
The metals deposited from SNe  are computed from a Chabrier IMF. The composition of each gas particle is traced throughout the simulation, being comprised of H, He and a range of metals  recycled from supernova explosions and deposited to the surrounding gas particles (see \citealt{stinson06}), with metal diffusion between gas particles also included \citep{shen10}.  \verb,GASOLINE, includes the effect of a uniform background radiation field on the ionization and excitation state of the gas.
Metal-line cooling \citep{shen10} and  feedback from massive stars \citep{stinson13} prior to their explosion as SNe are also included.

\begin{figure}
\hspace{0.cm}\includegraphics[width=8.5cm,height=12.2cm]{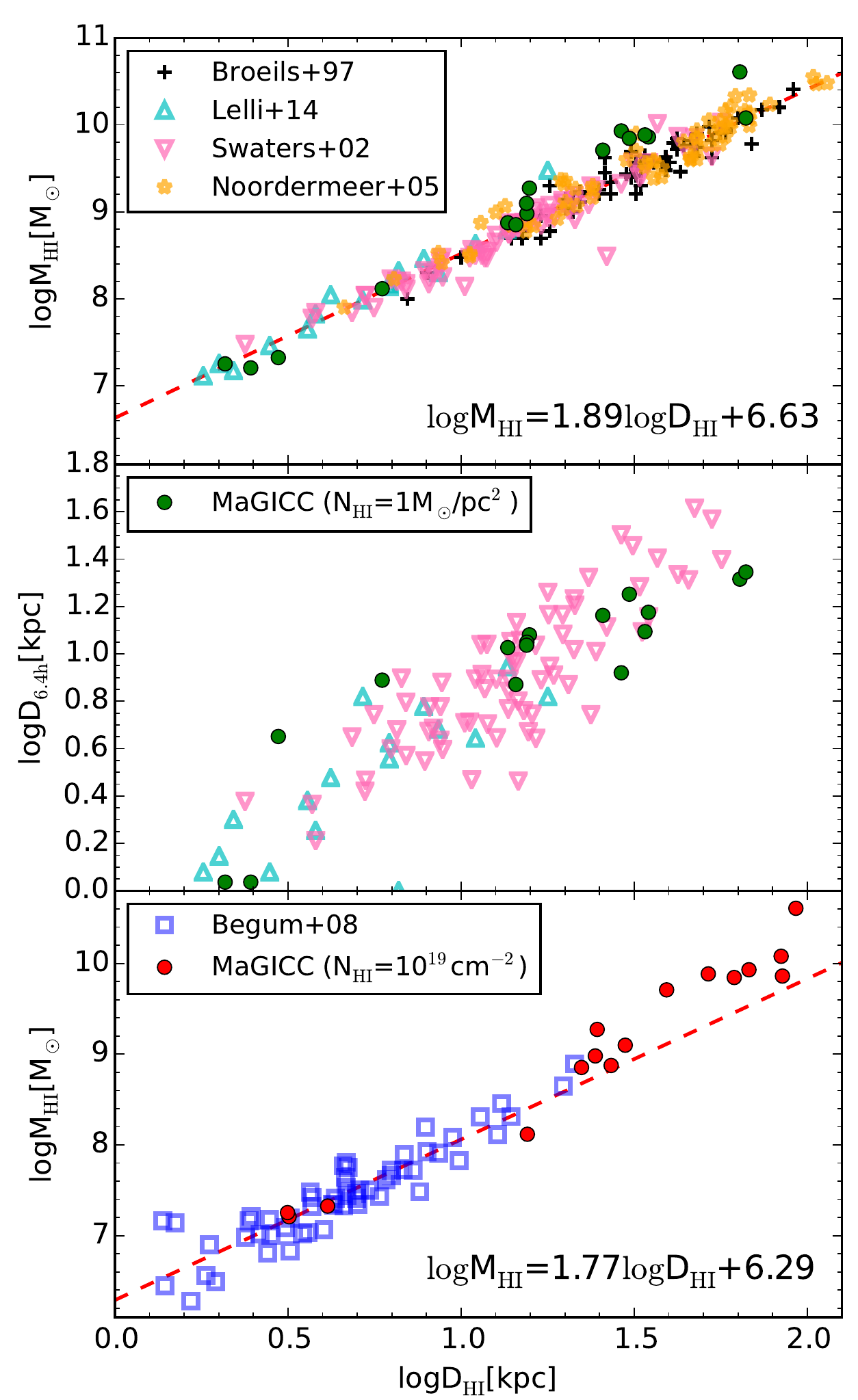}
\caption{Comparisons of the sizes of the simulated galaxies with observations, in HI gas and star light.  Observational data sets are shown in the legends.  
Top Panel: Green circles show \mhi of the simulations as a function of the diameter (\dHI) where the HI surface density drops below 
1\,\msun\,pc$^{-2}$. 
Middle panel:  \dsf\ as a function of \dHI\, where \dsf\ is 6.4 times the disc scale length.
 Bottom panel:  Red circles show \dHI\ where the HI surface density drops below  10$^{19}$ cm$^{-2}$ verses  \mhi. } 
\label{fig:logRHI_logMbar}
\end{figure}

 Radiative cooling has three parts $\Lambda$=$\Lambda_{H,He}$+$\Lambda_{metal}$+$\Lambda_{Comp}$.
where $\Lambda_{H,He}$ is net cooling due to (H, H$^+$, He, He$^+$ and He$^{++}$), $\Lambda_{metal}$ is the rate due to metals, and 
 $\Lambda_{Comp}$ is the Compton cooling and heating. For  $\Lambda_{H,He}$, cooling and heating rates are calculated directly from the ionization equations, meaning  the simulations capture the non-equilibrium cooling of primordial species. Each timestep takes the abundance of ions at the end of the previous step as the starting point.  Ionization, cooling and heating  rates match those of \cite{abel97}. For calculating HI fractions at z=0, we post process to include self-shielding, following Appendix 2 in \cite{rahmati13a} and ionization from star forming regions following \cite{rahmati13b}, using their Eqn 5 which relates star formation to gas surface density. The amount of HI is somewhat sensitive to the manner in which ionization rates are modeled, with neutral hydrogen masses around a factor of two larger using these models than when assuming that the ISM is optically thin.

There are 16 galaxies, separated into three  sub-sets,  labelled as Milky-Way (MW), irregular (Irr) and dwarf irregular (dIrr) types, with stellar masses ranging 5$\times$10$^{5}$-5$\times$10$^{10}$\msun.  Resolution varies depending on the ``type": MW types have m$_{\rm star}$=4.0$\times$10$^4$M$_{\odot}$, m$_{\rm gas}$=5.7$\times$10$^4$M$_{\odot}$, m$_{\rm dm}$=1.1$\times$10$^6$M$_{\odot}$ and a gravitational softening length of $\epsilon$=312pc (for all particle types);
 Irr's have m$_{\rm star}$=4.3$\times$10$^3$M$_{\odot}$, m$_{\rm gas}$=7.1$\times$10$^3$M$_{\odot}$, m$_{\rm dm}$=1.4$\times$10$^5$M$_{\odot}$ and $\epsilon$=156pc; and dIrr's have m$_{\rm star}$=5.7$\times$10$^2$M$_{\odot}$, m$_{\rm gas}$=1.1$\times$10$^3$M$_{\odot}$, m$_{\rm dm}$=1.7$\times$10$^4$M$_{\odot}$ and $\epsilon$=78pc.

The halos are identified using \ahf\ \citep{knollmann09} with halo masses  defined within a sphere containing $\Delta_{\rm vir}\simeq$350 times the cosmic matter density at  z=0. The bulk of the analysis is done using {\tt pynbody}   \citep{pynbody}.

For the simulations, we use baryon mass \mb=\mstar+4/3$\times$\mhi\, with the factor of 4/3 used to account for forms of gas other than HI, in a manner that mimics observational assumptions of the comparison data \citep[e.g.][]{mcgaugh12}.  \mhi\ remains an approximation since an accurate model of HI mass would require full radiative transfer. In particular self shielding from the UV background is not included. This uncertainty in the fraction of gas classified as HI will not affect our rotation curve shapes, but may affect the extent of the HI discs and hence the radius at which we measure \vrot. 
Disk scale lengths  are derived from exponential fits to the surface brightness profiles in the I band: each stellar particle from the simulation represents a single stellar population (SSP) for which an absolute magnitude in a particular bandpass is calculated, interpolating between a grid of SSP luminosities from  \cite{girardi10} and \cite{marigo08}  for various stellar ages and metallicities. We note that scale-lengths in Swaters et al. 2009 and Lelli et al. 2014 use the R band, which won't significantly effect the comparison.

Basic galaxy parameters are shown in Table~\ref{tab:sims}, including halo mass (\mhalo), stellar mass (\mstar), HI gas mass (\mhi), disc scale  length $h$, the extent of the HI disc \rHI, defined as the radius at which the HI density of the galaxy falls to 1\msun\,pc$^{-2}$, which is adopted in several observational studies \citep[eg.][]{broeils97,swaters02,noordermeer05,lelli14}, although we note that the \cite{begum08} data set which comprises a significant fraction of the observed  low mass galaxies, uses a threshold of 10$^{19}\,$cm$^{-2}$. We examine both thresholds in Fig~\ref{fig:logRHI_logMbar} which compares the sizes of the simulated galaxies with observations, in HI gas and star light. 
 Observational data come from \cite{broeils97,swaters02,noordermeer05,lelli14} and \cite{begum08}.
  
 Fig 1 plots \dHI\  against \mhi\ and \dsf. Green and red circles in the top and bottom panels show the simulations when defining the diameter (\dHI) as where the HI surface density drops below 1\,\msun/pc$^{-2}$  and 10$^{19}$ cm$^{-2}$ respectively. Fits to the observed \dHI-\mhi\ relation are shown as  red lines.
We will use this relation to define a radius \rHI= \dHI/2 where we measure the circular velocity \vlast, which results in a measure of \vrot\ for each simulation at a radius that is similar to that of observed galaxies of the same \mhi.
\dsf\ is defined as 6.4 times the disc scale length, and is used as an indication of the extent of star light in galaxies. This is considered better than using an isophotal diameter at e.g. 25 B-mag arcsec$^{-2}$, as only a small fraction of the disk is enclosed within the isophotal diameter in low surface brightness galaxies (Swaters et al. 2002).

The simulations fall reasonably within the range of observed galaxies, giving some confidence in the mass distributions of the baryons (see also \citealt{santos15}). Larger, uniform samples of simulations and observations are required to  make careful statistical comparisons including determination of scatter.  Regardless, the focus of this paper is on how different measures of \vrot\ give different BTFRs, with which we will proceed.

\begin{figure*}
\hspace{0.cm}\includegraphics[width=\textwidth]{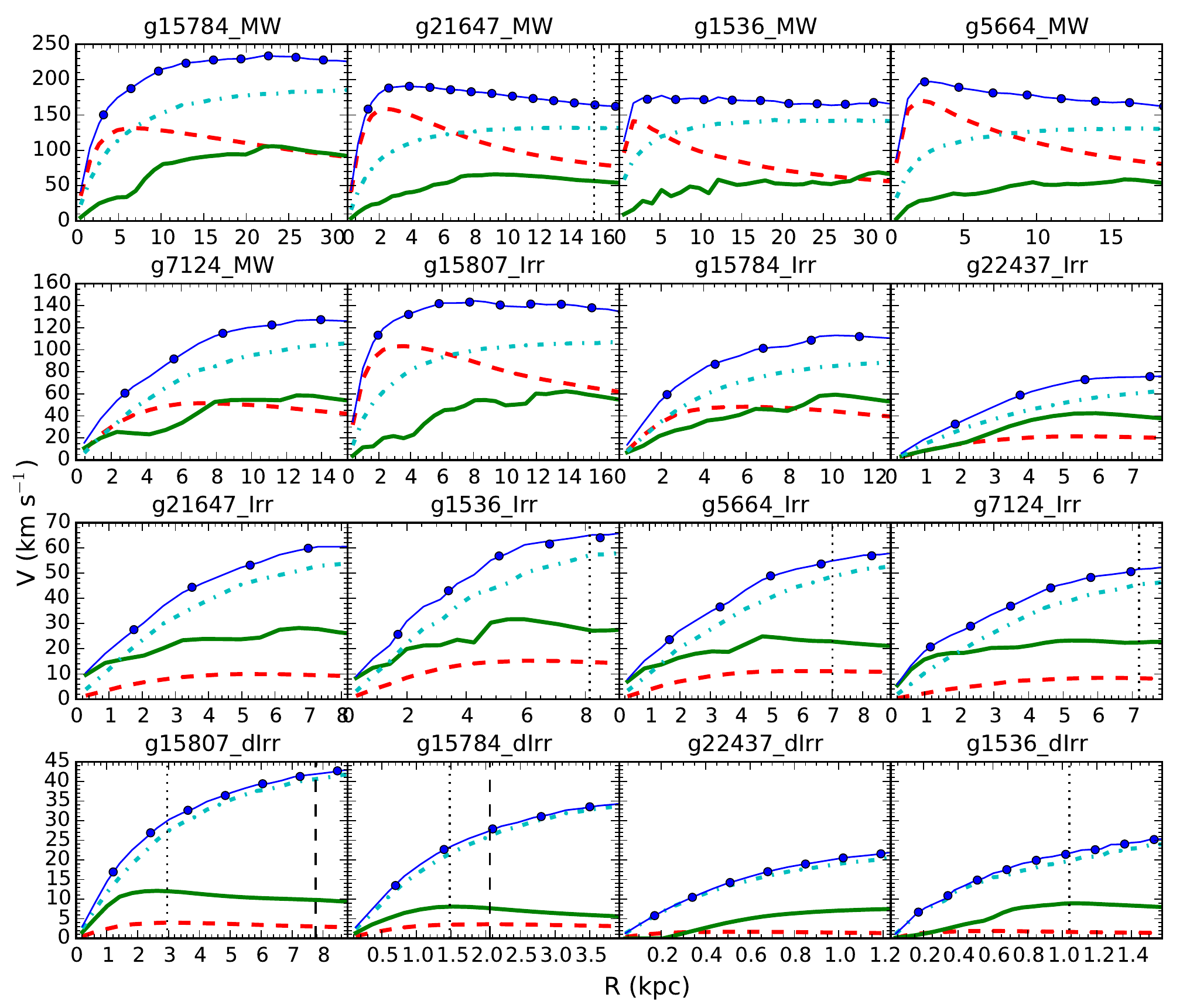}
\caption{Circular velocity rotation curves of the simulated galaxies, with different lines for the different components: solid for cold gas; dashed for stars; dot-dashed for dark matter; thin solid with circles for total. The circles in the total rotation curve mark multiples of $h$.  The curves extend to \rHI, or to \rflat\ in cases of rising rotation curves,  in which cases \rHI\ is marked as a dotted (dashed) vertical line when defined at 1\,\msun\,pc$^{-2}$  (10$^{19}\,$cm$^{-2}$).}
\label{fig:rc_magicc}
\end{figure*}


\section{Analysis and Results}\label{results}

 \subsection{Rotation curves}  \label{sec:rc}

Figure \ref{fig:rc_magicc} shows the circular velocity rotation curves of the simulated galaxies, computed using the  gravitational potential of the simulation  along the mid-plane of the  aligned disc. This is more accurate than just assuming a spherically symmetric potential and averaging mass inside spherical shells (i.e. the classical V$^2$=GM/r). The contribution  from each mass component is plotted as a different linestyle: cold gas thick solid; stars dashed; DM dot-dashed; total, thin solid with circles marking integer multiples of $h$. 
Rotation curves are plotted out to \rHI, or to \rflat, whichever is larger. Vertical dotted (dashed) lines show \rHI\ in cases where \rflat$>$\rHI, using density threshold of 1\,\msun\,pc$^{-2}$  (10$^{19}\,$cm$^{-2}$).

Whether the shapes of the rotation curves match, in detail, the distribution of observed galaxy rotation curve shapes requires a larger sample of simulations and is beyond the scope of this paper. We note that \cite{oman15} found that the EAGLE simulations \citep{schaye15} were not able to reproduce the variation in rotation curve shapes that is observed, in particular they could not simulate galaxies with slowly rising rotation curves. They showed that the MaGICC simulations, as used here, did form galaxies with such slowly rising rotation curves, as can be seen in Fig 2. Yet some observed low mass galaxies have more steeply rising rotation curves (e.g. Swaters et al. 2009). A larger suite of simulations will allow us to determine whether or not we can recover the full variation in rotation curve shapes that are seen in observations, which we will explore in a forthcoming study.

\subsection{HI linewidths} \label{sec:HI}
\begin{figure*}
\includegraphics[width=17.cm]{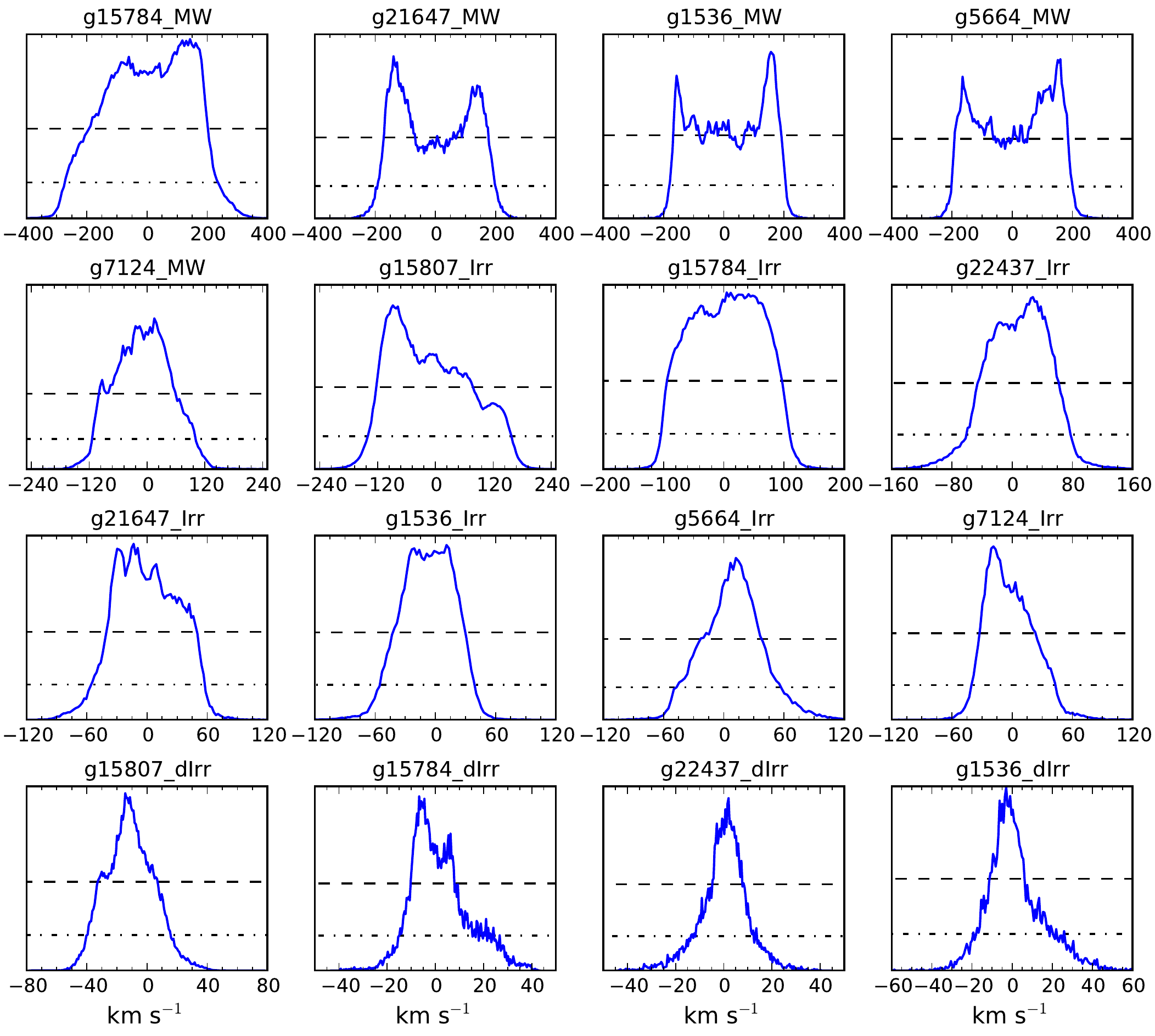}
\caption{HI line profiles of the MaGICC galaxies, shown for inclinations of 60$^\circ$ in each case.}
\label{fig:HIlineprof}
\end{figure*}

The observations of \vrot\ relevant to  this study are based on HI radio data, so we constructed mock HI data cubes.
We derive HI  line widths for the simulated galaxies, using inclinations 45$^\circ$, 60$^\circ$ and edge-on. In Fig~\ref{fig:HIlineprof}  we show the global HI line profiles for inclinations of 60$^\circ$, indicating 20\% and 50\% of the maximum flux by horizontal dashed and dot-dashed lines respectively. The quoted values of W$_{50}$ and W$_{20}$ in Table~\ref{tab:sims} are calculated by   taking the average of the inclination corrected line-widths from the different inclinations 45$^\circ$,  60$^\circ$ and edge-on.  The results we present do not change significantly if any single inclination is chosen.

\begin{figure}
\hspace{0.cm}\includegraphics[width=\columnwidth]{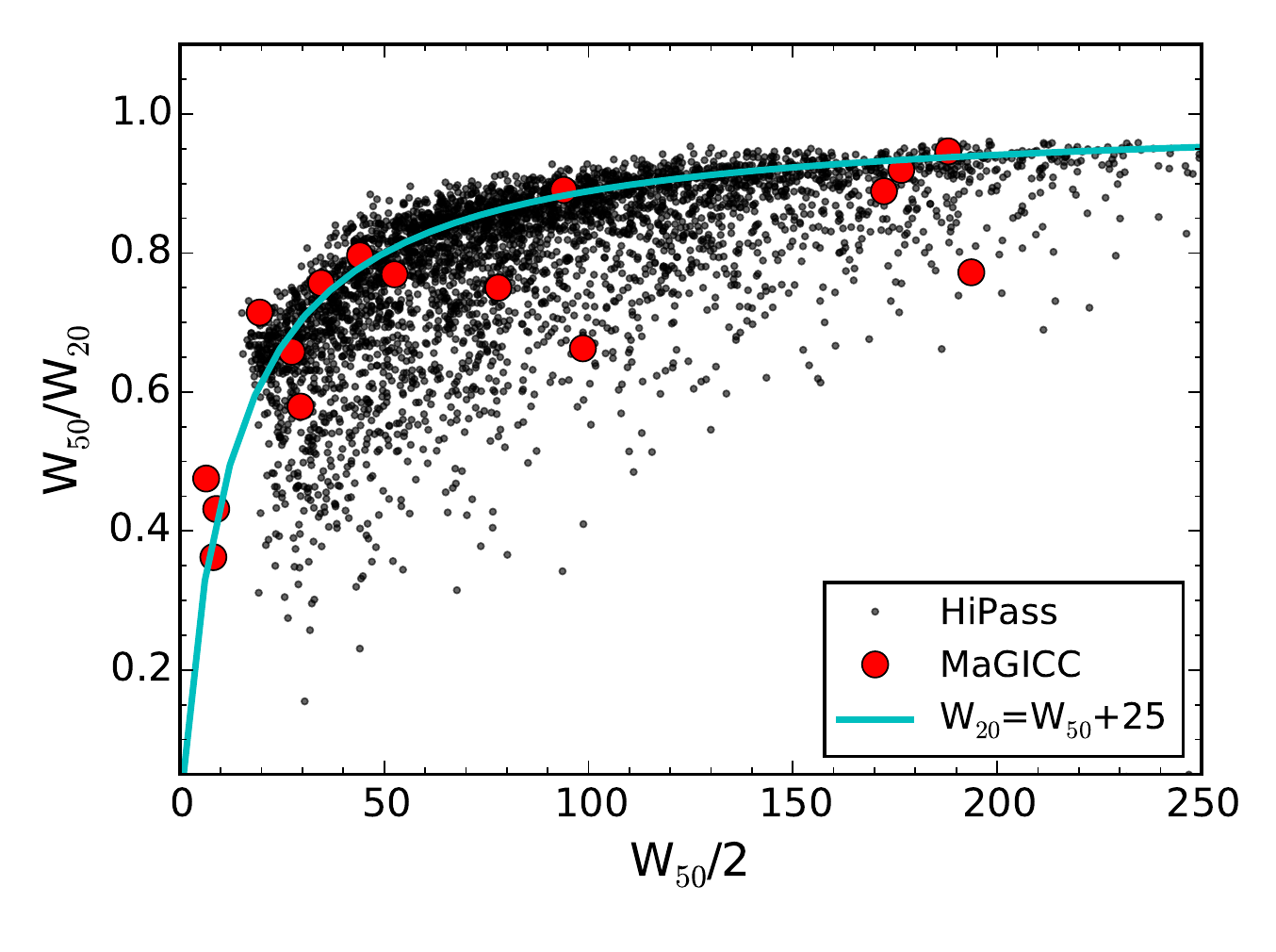}
\caption{The ratio \wf/\wt\ plotted as a function of \wt/2 for the simulations (red circles) overplayed on HiPass data (black dots), with the relation \wt=\wf$+$25. also shown as a cyan line.}
\label{fig:ratio}
\end{figure}

Fig~\ref{fig:ratio} shows the ratio \wf/\wt\  as a function of \wt/2   for the simulations (red circles) and HIPASS data (\citealt{meyer04} black dots). The  relation  \wt=\wf+25 is also shown  (cyan line), the typical difference between the observed \wt\ and \wf\ \citep{koribalski04,bradford15}. 
The offset between \wf\ and \wt\  becomes significant for lower velocities/masses, with the relation   \wf$\sim$\wt$-$25 implying that at \wt/2=[50,40,30], \wf/2 is  [25,31,42] per cent lower. Our simulations suggest that this simple relation holds down to very low velocities/masses.

\subsection{Measurements of rotation velocity}\label{sec:vrot}
We note some nomenclature  used in this paper:

\noindent$\bullet$ \vflat: the circular velocity at the flat part of the rotation curve.\\
$\bullet$ \rflat: the radius at which \vflat\ is reached.\\
$\bullet$ \vlast: the circular velocity at the last point of the HI disc.\\
$\bullet$ \rHI: the extent of the HI disc: defines where \vlast\ is measured.\\
$\bullet$ \dHI: diameter of the HI disc: 2$\times$ \rHI.\\ 
$\bullet$ \vmaxDM: the \vmax\ from the  DM only simulations.\\
$\bullet$ \rmaxDM: the radius at which \vmaxDM\ is reached.\\
$\bullet$ $h$: disc scale length.\\

\noindent The  criteria for selecting flat rotation curves in \cite{stark09}   requires that \vrot\ measured at 3 disc scale lengths (3$h$) is 85 per cent of  \vlast, or in cases where \rHI\ is greater than 4$h$, then \vrot\ measured at 4$h$ is required to be 90 per cent of \vlast. This definition  has a degree of arbitrariness, as it depends on the relative distance between \rHI\ and integer multiples of $h$. In this study, we define rotation curves to be flat in the regions where \vrot\ changes by less than 5 per cent between integer multiples of $h$. We define \vflat\  at the flattest of these flat regions, i.e. where the change in \vrot\ is the smallest, within the HI disc (in regions where radius$<$\rHI). \vflat\ is measured at the centre of the flattest region within the HI disc. 
 
 Our flatness criteria are better defined than those used in  \cite{stark09}, and are also stricter for galaxies with \rHI\ $<$5$h$, (69 per cent  of galaxies in the \citealt{swaters09} data set from which \citealt{stark09}  was largely drawn), is at least as strict for galaxies with 5$h$$<$\rHI$<$6$h$ (a further 9 per cent of  \citealt{swaters09} galaxies), but may not be as strict for galaxies with  \rHI\ $>$6$h$ (12 per cent of \citealt{swaters09} sample).

 Some of our low mass simulated galaxies do not have any flat regions at radii lower than \rHI. These are deemed to be galaxies with slowly rising rotation curves. In such cases, marked with $\dagger$ in Table~\ref{tab:sims}, we continue beyond \rHI\ until we find the first flat region of the rotation curve, i.e. a region where \vrot\ increases by less than 5 per cent as radius increases by $h$. \vflat\ is then measured at the central point of this flat region. 
 Galaxies with rising rotation curves will be  identified in our BTFR plot, and are analogous to those galaxies that are $excluded$ in studies of observed \vflat\ BTFRs \citep[e.g.][]{stark09,mcgaugh12,mcgaugh15}.

\subsection{The Baryonic Tully-Fisher relation}\label{sec:BTFR}
\begin{figure}
\hspace{0.cm}\includegraphics[width=\columnwidth]{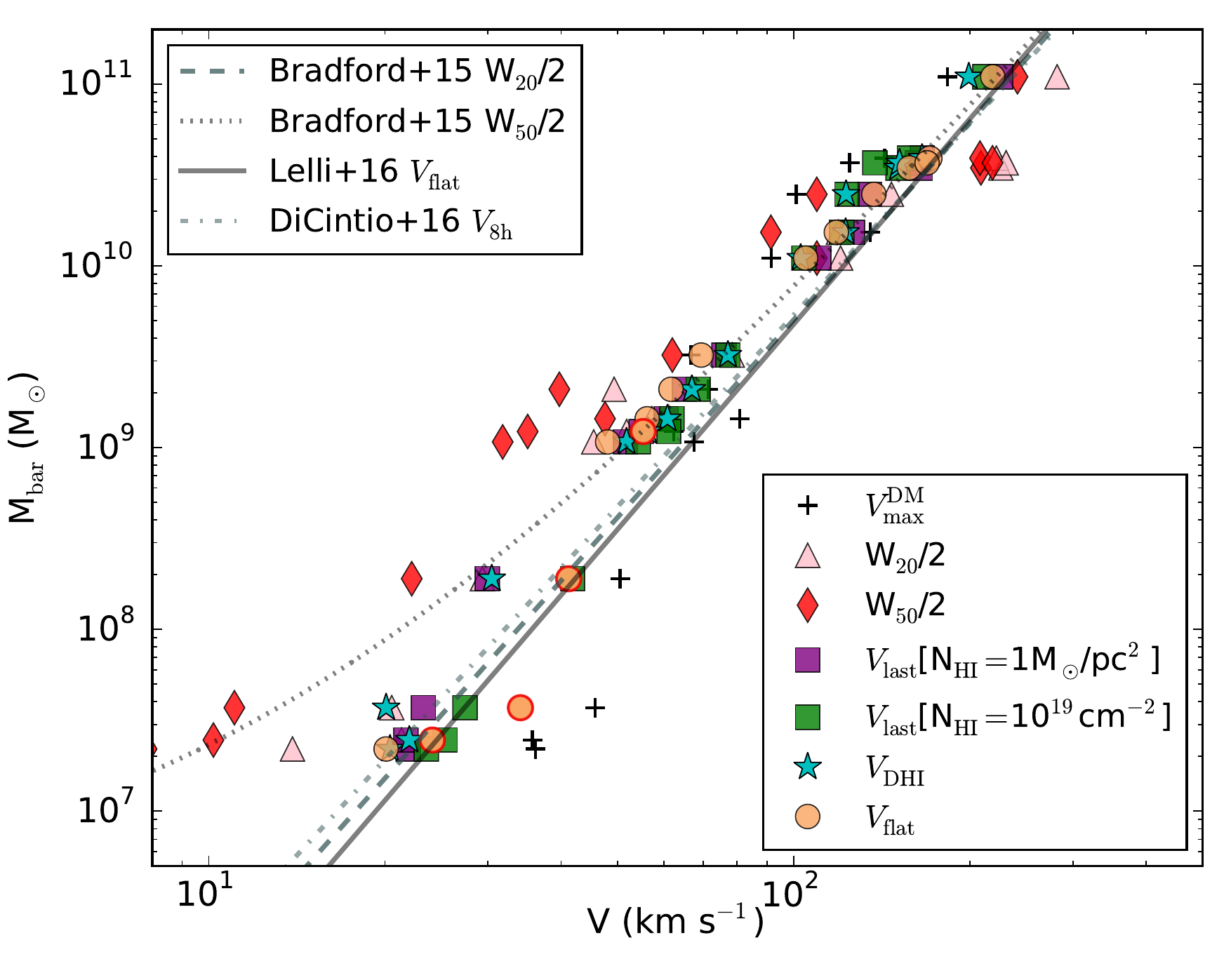}
\caption{BTFRs for the MaGICC galaxies. Different symbols show the relations obtained when using different \vrot\ measures:
 \vflat\ (orange circles), \vlast\ (green squares), \vDHI\ (blue stars),  \wt/2 (pink triangles), \wf/2 (red diamonds) and \vmaxDM\ (black crosses).
 The solid line is the observational BTFR obtained from gas-rich galaxies using  \vflat\ in  \citet{mcgaugh15}. The dashed line is the \wt/2 BTFR from \citet{bradford15}, while the dotted  line adjusts this \wt/2 relation using \wf=\wt$-$25, the typical difference between these values as found in \citet{bradford15}. The \vflat\ of  simulated galaxies with rising rotation curves at their outer point are indicated by a solid red outline: these  would be excluded from observed  \vflat\ BTFRs.}
\label{fig:btfr}
\end{figure}

Figure \ref{fig:btfr} shows the different BTFRs for the different \vrot\ measurements of our simulated suite of galaxies. These are:   \vflat\ (orange circles), \vlast\ defined using 1\,\msun/pc$^{-2}$  (purple squares) and 10$^{19}$ cm$^{-2}$  (green squares), \vDHI\ (blues stars) which uses the observed  \mb-\rlast\ relation to determine the radius to measure \vrot\ based on the \mb\ of each simulation,    \wt/2 (pink triangles), \wf/2 (red diamonds) and \vmaxDM\ (black crosses).

Overplotted are  observational BTFRs: the solid line shows the fit found in  \cite{lelli16}  using V$_{\rm flat}$ from observations of gas-rich galaxies log(\mb)=3.95log(\vflat)+1.86, a relation very similar to that found in \cite{mcgaugh15}; 
the dashed line shows the relation found by \citealt{bradford15} using \wt\ measurements of  galaxies selected from the Sloan Digital Sky Survey, log(\wt/2)= 0.277log(\mb)$-$0.672; 
the dotted line shows the \wf\ relation determined by using \wf=\wt$-$25. We found a similar \wf\ BTFR using local volume data as compiled by   \citealt{karachentsev13}, but do not overplot here as it adds to clutter without adding information. We also show a theoretical (dot-dashed) line from \cite{dicintio16}, which uses a semi-empirical model within a $\Lambda$ Cold Dark Matter cosmology, measured in the flat region of the rotation curve.

For simulated galaxies with \mb$\gsim$2$\times$10$^8$\msun\ the \vflat, \vlast\ and \wt\ BTFRs are  similar. This corresponds to galaxies with \vrot$\gsim$45\kms\ living in halos with 
\mhalo$\gsim$5$\times$10$^{10}$\msun.  \wt\ and \wf\ are integrated measurements that are  intrinsically more sensitive to contribution from the rising part of the rotation curves at small radii, which contains most of the HI flux, explaining their difference with \vlast. The difference \wf$\sim$\wt$-$25 holds over a large range of masses, meaning that the \wf\ BTFR does not have a large offset from the other measures of \vrot\ for massive galaxies: at \wt/2=[100, 150, 200], \wf/2 is  about [12,9,6] per cent lower than the other measures of \vrot.

The \vDHI\ BTFR differs from \vlast, indicative of the sensitivity of low mass galaxies to the radius at which \vrot\ is measured. This \vDHI\ BTFR also shows increased scatter at the the HI low mass end, but this is a region where scatter in the observed BTFRs also increases (see e.g. McGaugh 2012; Lelli et al. 2014).

For galaxies with \mb$\lsim$10$^8$\msun, the different forms of the BTFR increasingly diverge.  
The \vlast\ using threshold of  10$^{19}$ cm$^{-2}$ BTFR of the simulations follow closely the \vflat\ value of \cite{mcgaugh15}, which is also close to the \vflat\ BTFR from \cite{stark09}.  

The \vflat\ BTFR from the simulations also follows the observed  \vflat\ BTFR if the galaxies with rising rotation curves are neglected, as they are in the observational derivations. Such galaxies are marked with red outlines in Fig~\ref{fig:btfr}. Our model suggests that low mass galaxies with rising rotation curves would fall off the \vflat\ BTFR,  having excessively high values of \vrot, if their gas discs would extend far enough for \vflat\ to be measured.  

We also show results from the corresponding dark matter only simulations, which shows how the measured \vrot\ of disc galaxies becomes a smaller fraction of \vmaxDM. This is due to two related effects: primarily,  \rHI\ becomes a smaller fraction of \rmaxDM\ for low mass galaxies, which are confined to the  inner regions of DM halos (compare \rHI\ to \rmaxDM\  in Table~1, and see also Fig~7 of \citealt{brookshankar}); secondly,  baryons are able to cool  to the central regions of halos, so galaxies with high baryonic masses, i.e. high mass galaxies, have more steeply rising rotation curves in general, meaning that  \vrot\ becomes close to    \vmaxDM\ at relatively small radii.  

The shape of dark matter halos can be affected by baryons by either adiabatic contraction \citep{blumenthal86}, or  expansion (see \citealt{pontzen14}, for a review). The competing processes mean that the final profile can depend on the amount of gas inflowing to the central region causing contraction, and the timescale and mass of gas outflows causing expansion. Simulations suggest that this results in dark matter density profile shapes being dependent on galaxy mass \citep{DiCintio2014a,chan15,tollet16}. The rotation curve shapes in the galaxies of this study have been affected by these processes, but as stated above, the detailed study of these shapes is beyond the scope of this paper.

Bearing in mind that a larger number of simulations may be required to define a low mass BTFR, we nevertheless show in Table~\ref{tab2} the best fits to the various BTFRs from Fig. 5, with the form logM$_{\rm bar}$=$a$log\vrot+$b$. The difference in slopes between the various forms is significant.  We do note that linear fits cannot always be appropriate, for example the relation \wt=\wf$+$25 as shown in Fig. 4 implies that if the  \wt\ BTFR is well fit by a linear relation, then \wf\ would not be.

\section{Discussion}\label{discussion}

We have made various  measurements of \vrot\ for a suite of simulated galaxies within a \lcdm\ context, and derived the corresponding BTFRs.  Measuring \vrot\ using \vflat, \vlast, \vDHI, \wt, \wf\ and \vmaxDM\ result in significantly   different BTFRs, as  the different measures of \vrot\ are found to diverge at low masses. Each  BTFR derived using the different measures of \vrot\ from the simulations is consistent with the corresponding observed BTFRs in cases where observations are possible, i.e. \vlast,\vflat, \vDHI, \wt, \wf.

\begin{table}
\label{tab2}
\caption{Slopes ($a$) and zero points ($b$) of the various BTFRs}
\begin{center}
\begin{tabular}{c | c c }
\hline
\vrot & $a$ & $b$ \\
\hline
V$^{\rm DM}_{\rm max}$ & 5.19 & -.52 \\
W$_{20}$/2  &  3.13 & 3.78\\
W$_{50}$/2 & 2.70 &4.83\\
V$_{\rm last}$[N$_{\rm HI}=$1M$_\odot$\,pc${^-2}$] & 3.51 & 2.85\\
V$_{\rm last}$[N$_{\rm HI}=$10$^{19}$cm$^{-2}$] &  4.02 &  1.85\\
V$_{\rm DHI}$  & 3.56 & 2.80 \\
V$_{\rm flat}$ &  3.72 &  2.43 \\
\hline
\end{tabular}
\end{center}
\end{table}

Correspondence is found between simulations and observations in the \vflat\ case when  appropriate  selection of galaxies with flat rotation curves is made, excluding galaxies with rising rotation curves.  The simulations indicate  that galaxies with rising rotation curves would generally fall off the BTFR if their gas discs extended far enough to measure out to the flat part of the rotation curve.

We conclude that there is no single BTFR, with the relation between the baryonic mass and the rotation velocity of galaxies taking many forms which diverge significantly for low mass galaxies.  Differences in the BTFR for different measures of \vrot\  have been shown for observed galaxies \citep{verheijen01,noordermeer07,mcgaugh12,bradford16}. We show that the BTFR for simulated galaxies also depends on the manner in which rotation velocity is measured, with the  BTFRs becoming increasingly different from one another for low mass galaxies.
   
 The  various forms of the BTFR emerge from the fundamental processes in galaxy formation, relating to fundamental forces, mass acquisition,  angular momentum acquisition, baryon content,  gas cooling, star formation, and energy feedback. BTFRs are thus   crucial constraints on models that attempt to capture these fundamental processes of galaxy formation, not only in terms of understanding the relation between baryonic and total mass in galaxy populations, but also in terms of the velocity function \citep{brookshankar}, i.e, understanding the BTFR has important implications for how we compare the number of galaxies observed  as a function of \vrot, to the number predicted in an \lcdm\ cosmology. 

Our hydrodynamical simulations of galaxy formation  can explain the various forms of the BTFR within a  \lcdm\ context. 
 It is instructive to compare our study with two  recent  \lcdm\ cosmological models that have used the BTFR as a constraint, \cite{trujillogomez11} and \cite{dutton12}.

  In \cite{trujillogomez11}, a model population was made by matching  \mstar\ to \mhalo\ following abundance matching, and giving each model galaxy an  empirically motivated gas fraction and scale-length. A BTFR was derived  by measuring  circular velocities at 10\,kpc. For galaxies at the high mass end of our study, correspondence with  \cite{trujillogomez11} is good. This is expected because \rHI\ extends far enough  to reflect the value  of \vrot\ at 10\,kpc. For low mass galaxies, the  \cite{trujillogomez11} BTFR bends toward values of relatively high velocity, in a manner that is very similar to our \vmaxDM\ BTFR. This is also expected, because \rmaxDM\ is $\sim$10\,kpc for such galaxies, as seen in our Table~1, and such galaxies have very low baryon fractions meaning that DM only simulations are not greatly different from their model galaxies.  
   
 \cite{dutton12} use a semi-analytic model population of galaxies, tuned to match the empirical \mstar-\mhalo\ and specific angular momentum-\mhalo\ relations. They define \vflat\ as the circular velocity measured  at the radius that contains 80 per cent of the cold gas. This is analogous to our \vlast, so it is not surprising that both studies  match  the \vflat\ BTFRs from \cite{stark09}.  \cite{dutton12} does not test whether all their model galaxies' rotation curves are actually flat; the rotation curves in some low mass galaxies  may still be rising at the point they measure `\vflat'. In fact, correspondence with observations $requires$  the existence of low mass galaxies that have rising rotation curves, as was found in our simulations.

The rising rotation curves of many low mass galaxies means that one needs to take  particular care when comparing models and theory. Measuring the \vflat\ BTFR for low mass galaxies requires a selection, with galaxies with rising rotation curves discarded. It is not clear whether this introduces a bias. For example, do low mass galaxies with extended HI discs and flat rotation curves fit on the average \mstar-\mhalo\ relation? 

Measuring the \vlast\ BTFR, on the other hand,  is somewhat arbitrary, with the position of measurement determined by the radial  extent   of the cold gas. The large difference between \rHI\ and \rmaxDM\ highlights that \vrot\ may increase significantly in some cases, if HI discs extended further and allowed measurements at larger radii.  

In fact,  even when our low mass simulated  galaxies meet the criteria for \vflat, i.e. they rise by less than five per cent when radius increases by $h$,  they are still measured at a radius that can be a small fraction of \rmaxDM. In these cases, even small  gradients in the rotation curve can result in significant difference between the measured \vflat\ and the \vrot\ that would be reached if the cold gas extended far enough to measure rotation velocities at their peak value.

To proceed in confronting models with observations, it is suggested to test the BTFR as measured in different ways, at a variety of radii. This also facilitates the evaluation of the variation of rotation curve shapes, which provide further important constraints on galaxy formation models \citep{gentile07,mcgaugh07,oman15,brook2015}. Measuring at radii smaller than \rflat\ may also allow different cosmological models to be differentiated  \citep{brook15b}

Even better is to exploit the important information provided by the  entire rotation curve data of a large sample of galaxies. In a \lcdm\ model, one can  relate the halo mass as inferred  from the fit to the rotation curve  to   \mb, and compare this relation to the \mb-\mhalo\ relation that emerges from abundance matching studies. We will undertake such a procedure in a forthcoming study (Katz, H et al. in prep), comparing a \lcdm\ model that assumes an NFW profile \citep{navarro96} with a model where DM halo shapes are modified by baryons \citep{DiCintio2014b}.

\section*{Acknowledgements}
We thank the anonymous referee for helpful suggestions. CB thanks  
MINECO (Spain) for financial support through the grant AYA2012-31101 and  the Ramon y Cajal program. 
We thank DEISA for  access to  computing resources  through DECI projects SIMU-LU and SIMUGAL-LU and the allocation of resources
from STFC's DiRAC Facility (COSMOS: Galactic Archaeology),
the DEISA consortium, co-funded through EU FP6
project RI-031513 and the FP7 project RI-222919 (through the
DEISA Extreme Computing Initiative), the PRACE-2IP Project (FP7 RI-283493).

\bibliographystyle{mn2e}
\bibliography{archive}


\label{lastpage}

\end{document}